\documentclass[journal]{IEEEtran}

\ifCLASSINFOpdf
\else
   \usepackage[dvips]{graphicx}
\fi
\usepackage{url}

\usepackage{graphicx}
\usepackage{amsmath}
\usepackage{color}
\usepackage{amsfonts}
\usepackage{subfigure}
\usepackage{multirow}

\newcommand{\eg}{\textit{e}.\textit{g}.}

\begin{document}

\title{Parameter-Free Channel Attention for Image Classification and Super-Resolution}

\author{
Yuxuan Shi,
Lingxiao Yang,
Wangpeng An,
Xiantong Zhen,
Liuqing Wang
\thanks{
This work was supported in part by The National Natural Science Foundation of China (No. 62002176, 62176068, and 62171309). Corresponding author: Liuqing Wang (wlq@nankai.edu.cn)}
\thanks{Yuxuan Shi is with the School of Statistics and Data Science, Nankai University, Tianjin 300071, China. (E-mail: syxpop@outlook.com)
}
\thanks{Lingxiao Yang is with the School of Computer Science and Engineering, Sun Yat-sen University, Guangzhou 510006, China. (E-mail: yanglx9@mail.sysu.edu.cn)
}
\thanks{Wangpeng An is with Tiktok Inc, Mountain View, CA, USA, 94041. (E-mail: anwangpeng@gmail.com) .
}
\thanks{Xiantong Zhen is with United Imaging Healthcare, Co., Ltd., China (e-mail: zhenxt@gmail.com).
}
\thanks{Liuqing Wang is with Nankai University, Tianjin 300071, China. (E-mail: wlq@nankai.edu.cn)
}
}

\maketitle

\begin{abstract}
The channel attention mechanism is a useful technique widely employed in deep convolutional neural networks to boost the performance for image processing tasks, \eg, image classification and image super-resolution.
It is usually designed as a parameterized sub-network and embedded into the convolutional layers of the network to learn more powerful feature representations.
However, current channel attention induces more parameters and therefore leads to higher computational costs.
To deal with this issue, in this work, we propose a Parameter-Free Channel Attention (PFCA) module to boost the performance of popular image classification and image super-resolution networks, but completely sweep out the parameter growth of channel attention.
Experiments on CIFAR-100, ImageNet, and DIV2K validate that our PFCA module improves the performance of ResNet on image classification and improves the performance of MSRResNet on image super-resolution tasks, respectively, while bringing little growth of parameters and FLOPs.

\end{abstract}


\IEEEpeerreviewmaketitle

\section{Introduction}
\label{sec:intro}
\IEEEPARstart{A}{ttention} mechanisms~\cite{hu2018squeeze, dai2019second} have been validated to be very effective on boosting the performance of deep neural networks~\cite{hu2018squeeze,dai2019second,zhao2020efficient}.
Among them, channel attention~\cite{hu2018squeeze} is a commonly used attention mechanism that assign adaptive weights over different feature maps to improve the performance of image classification networks.
Therefore, it has become a key component of deep network architectures.
Besides, the effectiveness of channel attention~\cite{hu2018squeeze} on image super-resolution has also been illustrated in the work of deep residual channel attention network~\cite{zhang2018image} and cascading residual network~\cite{ahn2018fast}.

The parameter-free attention mechanism has been studied in~\cite{wang2018parameter} to assign position importance over the feature maps by spatial attention without parameter growth.
The parameter-free attention mechanism was also studied from the neuroscience perspective in~\cite{pmlr-v139-yang21o}. Different from current attention mechanism that needs to be trained from data, this work utilizes a fixed perception mechanism to pixel-wise importance over the feature maps without additional parameters.
Compared with popular parameterized channel or spatial attention mechanisms~\cite{hu2018squeeze,woo2018cbam}, parameter free attention mechanisms could achieve better effect~\cite{wang2018parameter,pmlr-v139-yang21o} while avoiding additional parameter growth, or reduce the model complexity and computational costs while preserving the network performance.

In this paper, we develop a Parameter-Free Channel Attention (PFCA) module to replace the standard channel attention module~\cite{hu2018squeeze} in popular image classification and image super-resolution networks. Our PFCA module exploits useful statistic information in different channels of a feature maps, and is a plug-and-play module can be embedded into a network to enhance its feature representation capability. By embedding our PFCA into ResNets and MSRResNet, experiments on CIFAR-100, ImageNet, and DIV2K demonstrate that our PFCA module is effective on boosting the performance of ResNet-18, ResNet-50, ResNet-101 and MSRResNet on image classification and super-resolution, respectively, while bringing little computational costs and parameter growth.

The rest of this paper is organized as follows.
In Section \ref{sec:related}, we briefly introduce the related works.
In Section \ref{sec:method}, we present our PFCA module for image classification and super-resolution tasks.
Extensive experiments are conducted in Section \ref{sec:experiment} to evaluate the performance of our PFCA module.
Section \ref{sec:conclusion} concludes this work.

\section{Related Works}
\label{sec:related}

\subsection{Image Classification}

Image classification has been widely tackled by convolutional neural networks (CNNs) ever since AlexNet~\cite{krizhevsky2012imagenet}.
Then VGG~\cite{simonyan2014very} extends the network's depth and width.
GoogleNet~\cite{szegedy2015going} provides an inception module to reduce the parameters of convolutional networks. The design of CNN architecture could affect the feature extraction function. In this work, we use deep residual network (ResNet)~\cite{he2016deep}, the most frequently used CNN architecture,  as the backbone for image classification tasks.
It has been validated by squeeze and excitation network (SE-Net)~\cite{hu2018squeeze} that attention mechanism helps improve the effectiveness of network to concentrate on special channels of feature maps. Improvements in optimization take effect to boost the network training process~\cite{an2018pid}. The study of feature representation has also been testified to be useful for classification~\cite{xu2019sparse}.


\begin{figure}[t!]
	\centering
    \small
	\includegraphics[width=\linewidth]{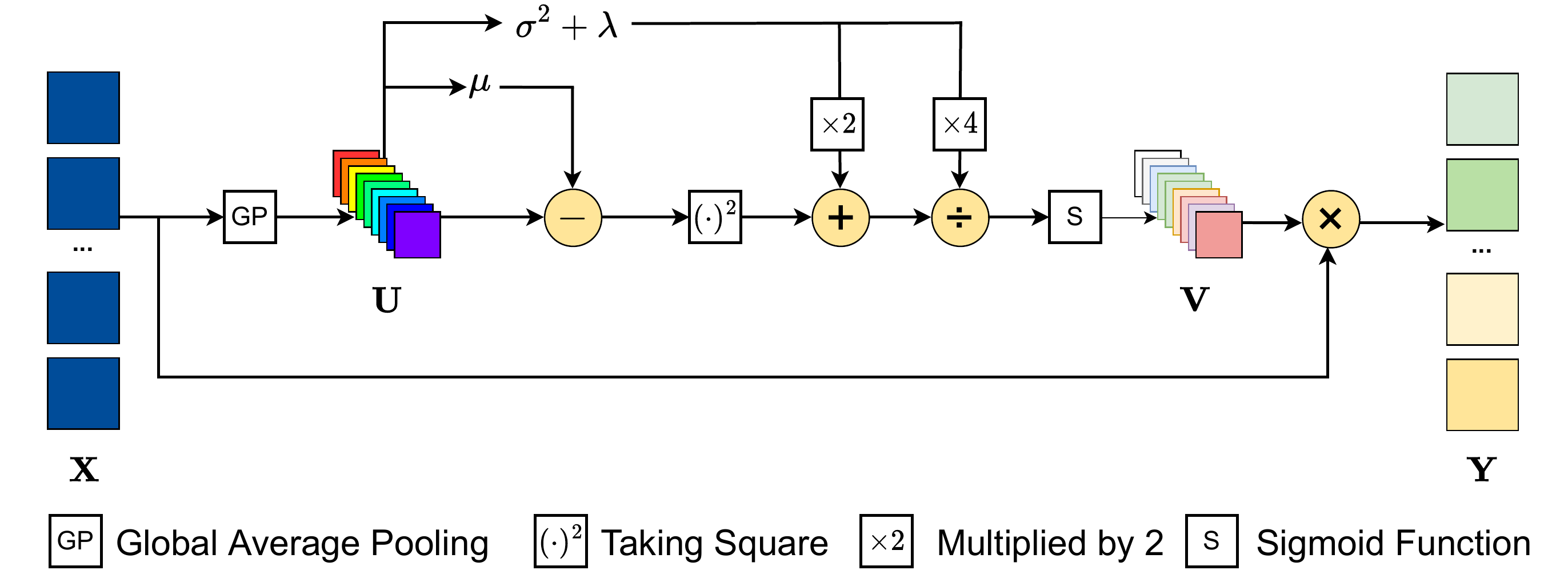}
	\vspace{-6mm}
	\caption{Illustration of our Parameter-Free Channel Attention (PFCA) module. The module gets the feature map $\mathbf{X}$ as the input and outputs feature map $\mathbf{Y}$. $\mathbf{U}$ and $\mathbf{V}$ are the channel level vector generated after the global pooling layer and sigmoid activation respectively. $\mu$ and $\sigma^2$ are the channel-level mean and variance of $\mathbf{U}$ and $\lambda$ is used to control variance.
	%
	}
    \label{fig:pcafreemodule}
\end{figure}

\subsection{Image Super-Resolution Task}

Using convolutional neural network to realize image super-resolution (SR) was firstly proposed in SRCNN~\cite{dong2015image}. The direct use of original LR image in the following ESPCNN~\cite{shi2016real} as input of the network instead of the upsampled image in SRCNN is to reduce the computation load. After that, VDSR~\cite{kim2016accurate} introduces the concept of residual image super-resolution , which has been proved to be effective in SR tasks. Since then, the residual block was provided in~\cite{ledig2017photo} to help design the network architecture of SR network, which was generally used in many following works. 
However, these methods could hardly be applied to practical scenarios due to their heavy computations, thus pushing the efficient SR network coming out.
Some methods are presented including depth-wise convolutions~\cite{huang2015single} and group convolutions~\cite{bevilacqua2012low,matsui2017sketch,he2019modulating} in order to save computation. The attention mechanism is also used to improve SR performance. RCAN~\cite{zhang2018image} uses channel attention module to make its network concentrate on important channels. PAN~\cite{zhao2020efficient} introduces a pixel attention module to adaptively weight different pixels of feature maps.  

\subsection{Attention Mechanisms}
Attention mechanisms have attracted great attention since it was successfully applied in natural language processing tasks~\cite{bahdanau2014neural} , which is mainly used to help networks concentrating on partial information of data flow and generating the corresponding dimensional weight maps.
When it comes to the image processing task, there are mainly three types of attention mechanisms: channel attention (CA)~\cite{hu2018squeeze}, spatial attention (SA)~\cite{dai2019second} and pixel attention (PA)~\cite{zhao2020efficient}, depending on their generated weight maps' dimension level.
Self-Attention~\cite{vaswani2017attention} is also very popular apart from the three.
%
Among the above mechanisms, channel attention is the first to propose in SE-Net~\cite{hu2018squeeze} and then polished in SK-Net~\cite{li2019selective}. After that, attempts to incorporate both channel attention and spatial attention for stronger attention capability in Convolutional Block Attention Module (CBAM)~ \cite{woo2018cbam} are tried several times. 
Bottleneck Attention module (BAM)~\cite{park2018bam} also uses double attention modules to add their attention matrices to get the final attention map.
SA-Net~\cite{zhang2021sa} splits the channel feature maps separately into the channel attention module and spatial attention module. Apart from these, there also exists the temporal attention, i.e. the attention in time level~\cite{xu2021temporal}.

\section{Parameter-Free Channel Attention}
\label{sec:method}

\subsection{Preliminary on Channel Attention Mechanism}

To demonstrate the Parameter-Free Channel Attention module (PFCA), we should first review the channel attention mechanism. Consider the input feature map $\mathbf{X}\in \mathbb{R}^{N\times C\times H\times W}$, where $N$ is batch, $C$ is channel and $H\times W$ is feature map size, the channel attention module is to compute the weighted map from the channel-level vector obtained by a pooling layer on $\mathbf{X}$, such as the average pooling layer. The squeeze and excitation process~\cite{hu2018squeeze}, including a well-known module which consists of two parameterized Multilayer Perceptrons (MLP), helps to integrate the channel level information and outputs the weighted map. The weighted attention map is multiplied by input $\mathbf{X}$ to get the final output $\mathbf{Y}$.
This process can be described as
\begin{equation}
\setlength\abovedisplayskip{2pt}
\setlength\belowdisplayskip{2pt}
    \mathbf{Y} = \mathbf{X} \cdot \text{Attention}(\text{Pooling}(\mathbf{X}))
\end{equation}


Channel attention is like a gate that controls the flow of feature maps in the channel level, making some of these feature maps more important to be processed by the inter network module. In other words, this two MLP parts could function as an ``importance" interpreter for feature maps. In this work, we try to provide a method that is similar to  channel attention but does not include additional parameters.

\subsection{Parameter-Free Channel Attention}
The Parameter-Free Channel Attention (PFCA) is similar to the channel attention of the channel attention (CA) mechanism. PFCA is inspired by simple attention module~\cite{pmlr-v139-yang21o} which aims to provide a parameter-free pixel-level attention. We apply this idea to the channel-level attention. The attention map $\mathbf{V}=(\mathbf{V}_1, \mathbf{V}_2, ...,\mathbf{V}_C)$ $\in$ $\mathbb{R}^{N\times C\times1\times1}$ is attained from its channel-level vector $\mathbf{U}=(\mathbf{U}_1, \mathbf{U}_2, ...,\mathbf{U}_C)$ $\in$ $\mathbb{R}^{N\times C\times1\times1} $ which is also provided by a pooling layer, e.g. average pooling layer. Then the channel-level vector $\mathbf{U}$ is computed by its element-wise operation below: 
\begin{equation}
\setlength\abovedisplayskip{2pt}
\setlength\belowdisplayskip{2pt}
    \mathbf{V}_j = \frac{ ( \mathbf{U}_j - \mu)^2 + 2(\sigma^2+ \lambda)}{4(\sigma^2+ \lambda) }~,~j=1,2,...,C
\end{equation}

Here $\mu$ and $\sigma^2$ are respectively the mean value and variance of channel vector $\mathbf{U}$ computed in channel dimension. $\lambda$ is a small value to control the variance and is set to 1e-4 in the following tasks. After that, the weighted attention map $\mathbf{V}$ is generated by the sigmoid activation of the above part and then is multiplied by input $\mathbf{X}$ to get the output result.
\begin{equation}
\setlength\abovedisplayskip{2pt}
\setlength\belowdisplayskip{2pt}
    \mathbf{Y} = \mathbf{X}\cdot \text{sigmoid}(\mathbf{V})
\end{equation}
The overall PFCA module is illustrated in Figure~\ref{fig:pcafreemodule}.
The pooling layer and the activation part is the same with channel attention. 
We perform a fixed perception to analyze the channel importance.

\begin{figure}[t!]
	\centering
    \small
	\includegraphics[width=\linewidth]{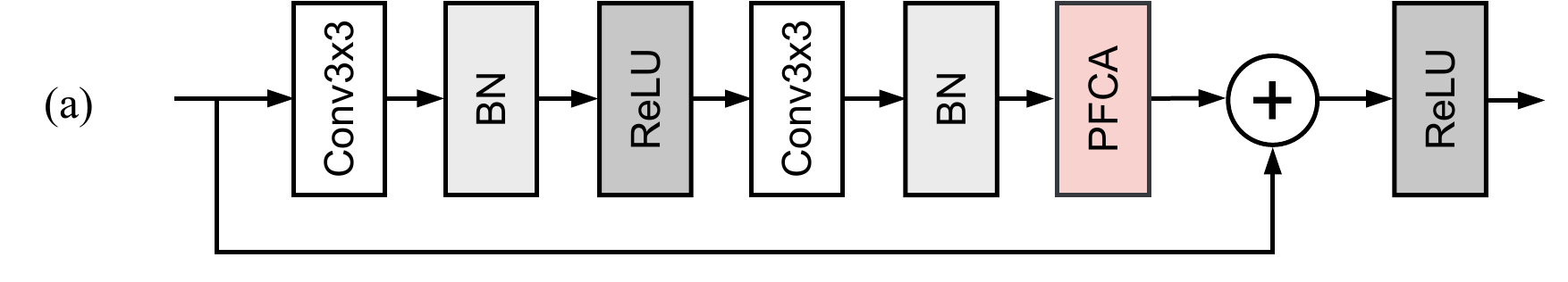}
	\includegraphics[width=\linewidth]{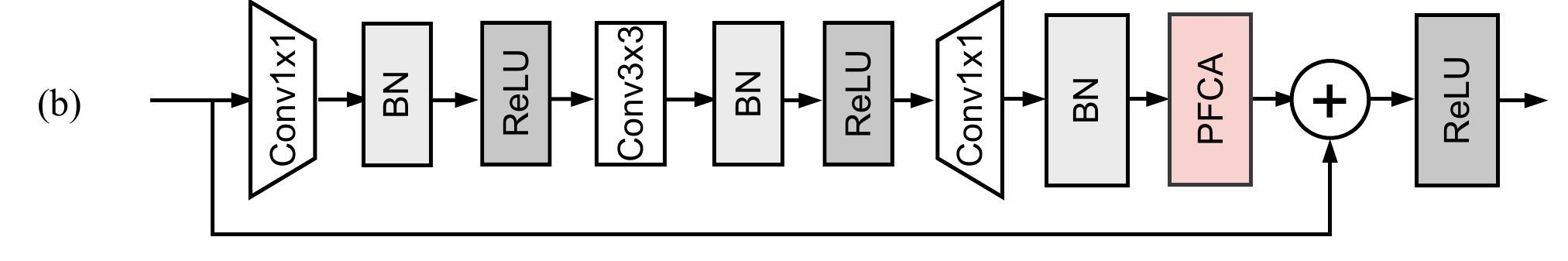}
	\vspace{-6mm}
	\caption{Illustration of the basic block~(a) and the bottleneck block~(b) with PFCA module used in our image classification model. Both PFCA are inserted before the addition operation.
	%
	}
    \label{fig:pfca-classification}
\end{figure}

\begin{figure}[h]
	\centering
    \small
	\includegraphics[width=\linewidth]{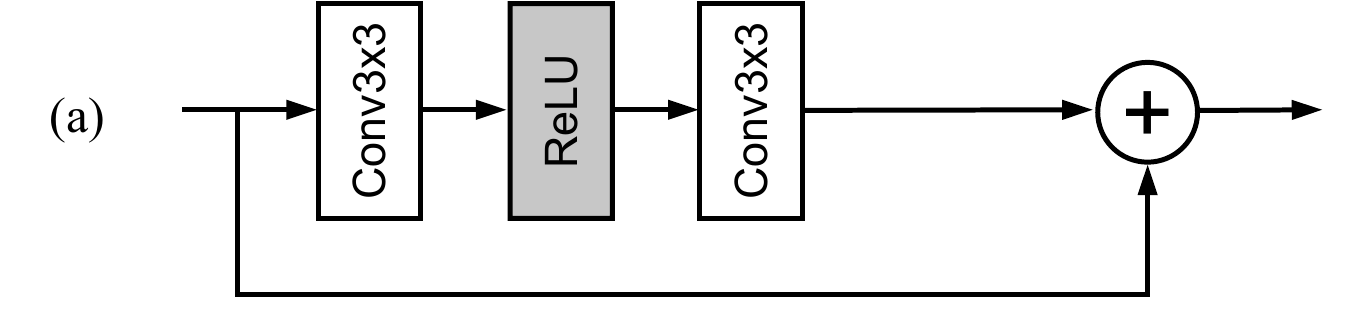}
	\includegraphics[width=\linewidth]{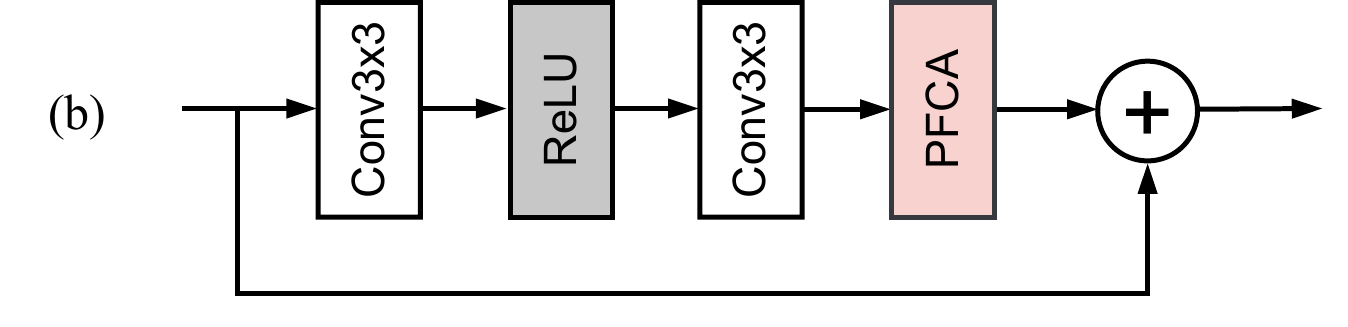}
	\vspace{-6mm}
	\caption{Illustration of the modified compact residual block~(a), and the proposed residual block with PFCA module~(b) in our image super-resolution model. PFCA is inserted after the second convolution layer and before the addition operation in each compact residual block.
	%
	}
    \label{fig:pfca-sr}
\end{figure}

\subsection{Image Classification and Super-Resolution Networks}

Our PFCA module is designed to make a difference in raising the current network performance, so we could insert the PFCA module into commonly used CNN models. However, the network structures would be different when inserting the proposed module in the object models since they vary in different tasks. To make the PFCA module more applicable, here we choose the widely used residual blocks~\cite{he2016deep} to build our image classification and image super-resolution models.


\noindent
\textbf{Image classification}.
Here, we use the ResNet-18/50/101~\cite{he2016deep} as our image classification backbones.
We first embed our PFCA module into the basic residual block in ResNet-18, and into the bottleneck residual block in ResNet-50 and ResNet-101~\cite{he2016deep}. The corresponding revised residual block and bottleneck residual block are illustrated in Figure~\ref{fig:pfca-classification}. The PFCA module is put before the addition operation in the basic residual block and the bottleneck residual block~\cite{he2016deep}, similar to~\cite{hu2018squeeze}. Then we replace the basic residual block in ResNet-18 by our revised residual block, and replace the bottleneck residual block in ResNet-50 and ResNet-101 by our revised bottleneck residual block. 


\noindent
\textbf{Image super-resolution}.
Here, we employ the MSRResNet~\cite{DBLP:conf/eccv/ZhangDLTLTWZHXL20} as the baseline super-resolution model, which uses the SRResNet~\cite{ledig2017photo} as the backbone but remove all the Batch Normalization blocks~\cite{ioffe2015batch}. The baseline model MSRResNet utilizes a compact residual block instead of the original one in~\cite{ledig2017photo}. Then we put the PFCA module between the second convolution layer and the addition operation in the compact residual block, as shown in Figure~\ref{fig:pfca-sr}. 
We replace the compact residual block by our revised one in MSRResNet.

\begin{table}[h]
\vspace{-5mm}
\caption{Classification accuracy (\%) of different methods on ImageNet~\cite{krizhevsky2012imagenet} valid set and CIFAR-100~\cite{krizhevsky2009learning} test set.
The number of parameters (M) and FLOPs (G) are reported.}
\begin{center}
\begin{tabular}{|l|cc|c|cc|}
\hline
\multirow{2}{*}{\small Model} & \multirow{2}{*}{\small Params} & \multirow{2}{*}{\small FLOPs} & \multirow{2}{*}{\small CIFAR-100} & \multicolumn{2}{c|}{\small ImageNet}  \\ \cline{5-6} 
&&                           &                               & \multicolumn{1}{c}{\small TOP 1} & \small TOP 5 \\ \hline
ResNet18               & 11.15                     & 1.69                     & 75.71                         & \multicolumn{1}{c}{70.49} & 89.55 \\
+ CA                   & 11.23                     & 1.70                     & 76.22                         & \multicolumn{1}{c}{71.05} & 90.15 \\
+ PFCA                & 11.15                     & 1.70                     & 75.93                         & \multicolumn{1}{c}{70.65} & 89.76 \\ \hline
ResNet50               & 24.37                     & 3.83                     & 78.66                         & \multicolumn{1}{c}{75.75} & 92.57 \\
+ CA                   & 26.79                     & 3.84                     & 78.85                         & \multicolumn{1}{c}{76.97} & 93.55 \\
+ PFCA                & 24.37                     & 3.83                     & 79.00                         & \multicolumn{1}{c}{76.82} & 93.44 \\ \hline
ResNet101              & 42.49                     & 7.30                     & 78.79                         & \multicolumn{1}{c}{77.52} & 93.73 \\
+ CA                   & 47.04                     & 7.31                     & 79.56                         & \multicolumn{1}{c}{78.02} & 94.12 \\
+ PFCA                & 42.49                     & 7.30                     & 79.11                         & \multicolumn{1}{c}{77.98} & 93.91 \\ \hline
\end{tabular}
\end{center}
\vspace{-5mm}
\label{table-imagenet}
\end{table}

\begin{table*} [!hbt]
\caption{\textbf{PSNR/SSIM results of comparison image super-resolution methods}.
``PA'', ``CA'', and ``PFCA'' means Pixel Attention, Channel Attention, and Parameter-Free Channel Attention, respectively.
FLOPs are computed on a $256\times256$ image.}
\begin{center}
\vspace{-3mm}
\begin{tabular}{|l|r|r|c|c|c|c|c|}
\hline
Model  & Params (K) & Flops (G) & Set5           & Set14          & B100           & Urban100       & Manga109       \\ \hline
SRCNN~\cite{dong2015image} & 20.1 & 21.07 & 30.48/0.8628 & 27.49/0.7503 & 26.90/0.7101 & 24.52/0.7221 & 27.66/0.8505 \\
EDSR~\cite{lim2017enhanced} & 43089.9  & 3293.90 & 32.48/0.8988 & 28.81/0.7879 & 27.72/0.7420 & 26.65/0.8036 & 31.04/0.9158
\\
\hline
MSRResNet~\cite{DBLP:conf/eccv/ZhangDLTLTWZHXL20} & 1517.6     & 166.36    & 32.19/0.8954 & 28.65/0.7835 & 27.60/0.7374 & 26.20/0.7900 & 30.55/0.9101 \\
+ PA     & 1584.1     & 170.66    & 32.24/0.8956 & 28.64/0.7834 & 27.61/0.7375 & 26.19/0.7900 & 30.55/0.9099 \\
+ CA     & 1526.9     & 166.36    & 32.26/0.8963 & 28.65/0.7829 & 27.60/0.7373 & 26.26/0.7914 & 30.70/0.9115 \\
+ PFCA  & 1517.6     & 166.36    & 32.26/0.8961 & 28.69/0.7838 & 27.61/0.7375 & 26.26/0.7917 & 30.69/0.9115 \\ \hline
\end{tabular}
\end{center}
\vspace{-6mm}

\label{table-sr-evaluation}
\end{table*}


\begin{figure*}[htb]
\vspace{-0mm}
\centering
\subfigure{
\begin{minipage}[t]{0.15\textwidth}
\centering 
\raisebox{-0.15cm}{\includegraphics[width=1\textwidth]{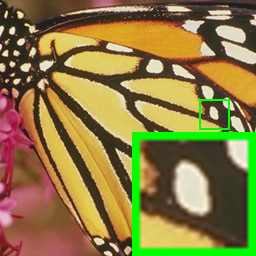} }
{\footnotesize PSNR/SSIM}
\end{minipage}
\begin{minipage}[t]{0.15\textwidth}
\centering 
\raisebox{-0.15cm}{\includegraphics[width=1\textwidth]{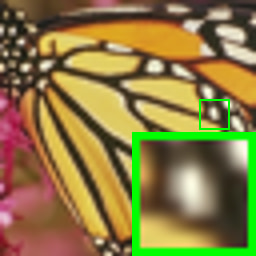}}
{\footnotesize 22.32 dB/0.7379 }
\end{minipage}
\begin{minipage}[t]{0.15\textwidth}
\centering 
\raisebox{-0.15cm}{\includegraphics[width=1\textwidth]{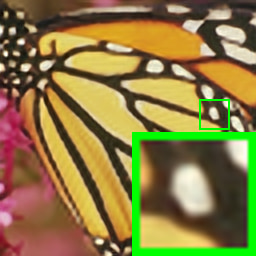}}
{\footnotesize 25.03 dB/0.8494}
\end{minipage}
\begin{minipage}[t]{0.15\textwidth}
\centering 
\raisebox{-0.15cm}{\includegraphics[width=1\textwidth]{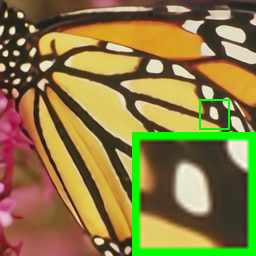}}
{\footnotesize 29.10 dB/0.9306}
\end{minipage}
\begin{minipage}[t]{0.15\textwidth}
\centering 
\raisebox{-0.15cm}{\includegraphics[width=1\textwidth]{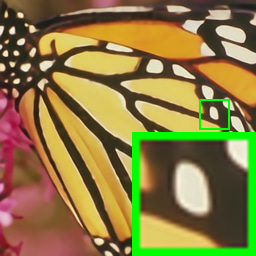}}
{\footnotesize 28.61 dB/0.9257}
\end{minipage}
\begin{minipage}[t]{0.15\textwidth}
\centering 
\raisebox{-0.15cm}{\includegraphics[width=1\textwidth]{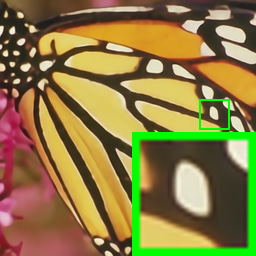}}
{\footnotesize 28.72 dB/0.9267}
\end{minipage}
}\vspace{0mm}
\subfigure{
\begin{minipage}[t]{0.15\textwidth}
\centering 
\raisebox{-0.15cm}{\includegraphics[width=1\textwidth]{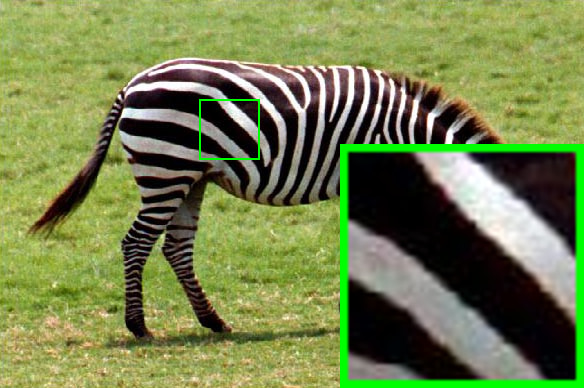} }
{\footnotesize  PSNR/SSIM }
\end{minipage}
\begin{minipage}[t]{0.15\textwidth}
\centering 
\raisebox{-0.15cm}{\includegraphics[width=1\textwidth]{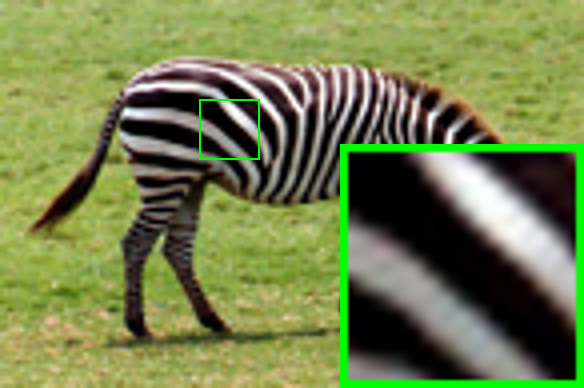}}
{\footnotesize  24.34 dB/0.6946 }
\end{minipage}
\begin{minipage}[t]{0.15\textwidth}
\centering 
\raisebox{-0.15cm}{\includegraphics[width=1\textwidth]{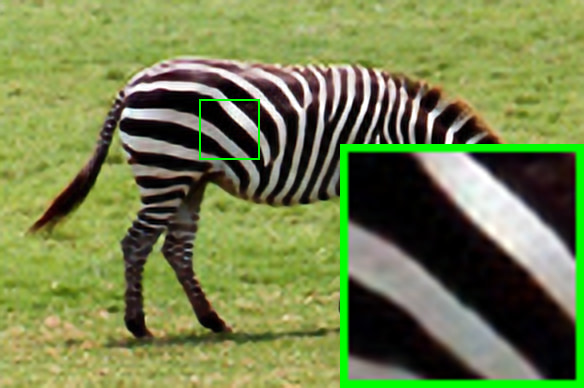}}
{\footnotesize 26.02 dB/0.7504 }
\end{minipage}
\begin{minipage}[t]{0.15\textwidth}
\centering 
\raisebox{-0.15cm}{\includegraphics[width=1\textwidth]{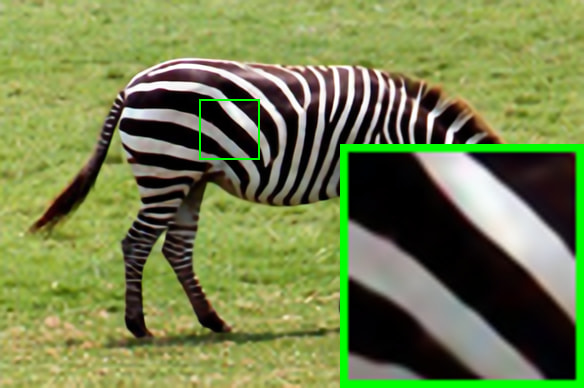}}
{\footnotesize  27.95 dB/0.7901 }
\end{minipage}
\begin{minipage}[t]{0.15\textwidth}
\centering  
\raisebox{-0.15cm}{\includegraphics[width=1\textwidth]{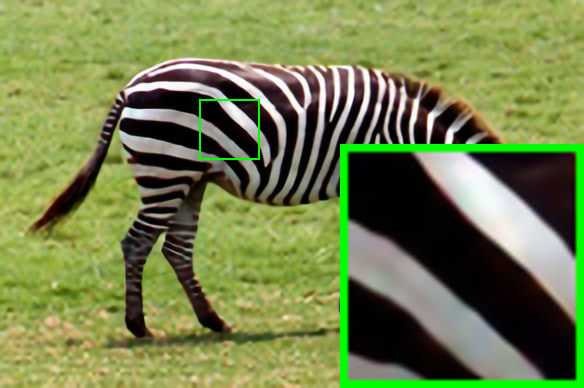}}
{\footnotesize  27.52 dB/0.7830 }
\end{minipage}
\begin{minipage}[t]{0.15\textwidth}
\centering 
\raisebox{-0.15cm}{\includegraphics[width=1\textwidth]{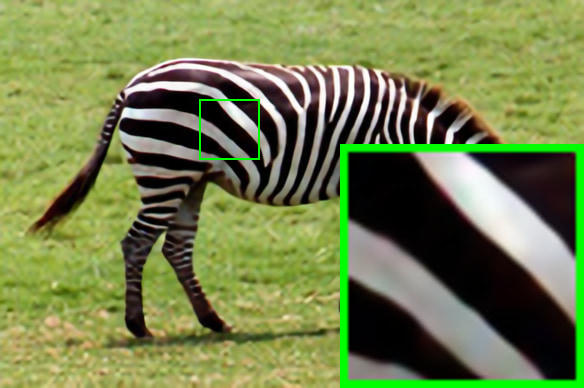}}
{\footnotesize 27.61 dB/0.7854 }
\end{minipage}
}
\subfigure{
\begin{minipage}[t]{0.15\textwidth}
\centering 
\raisebox{-0.15cm}{\includegraphics[width=1\textwidth]{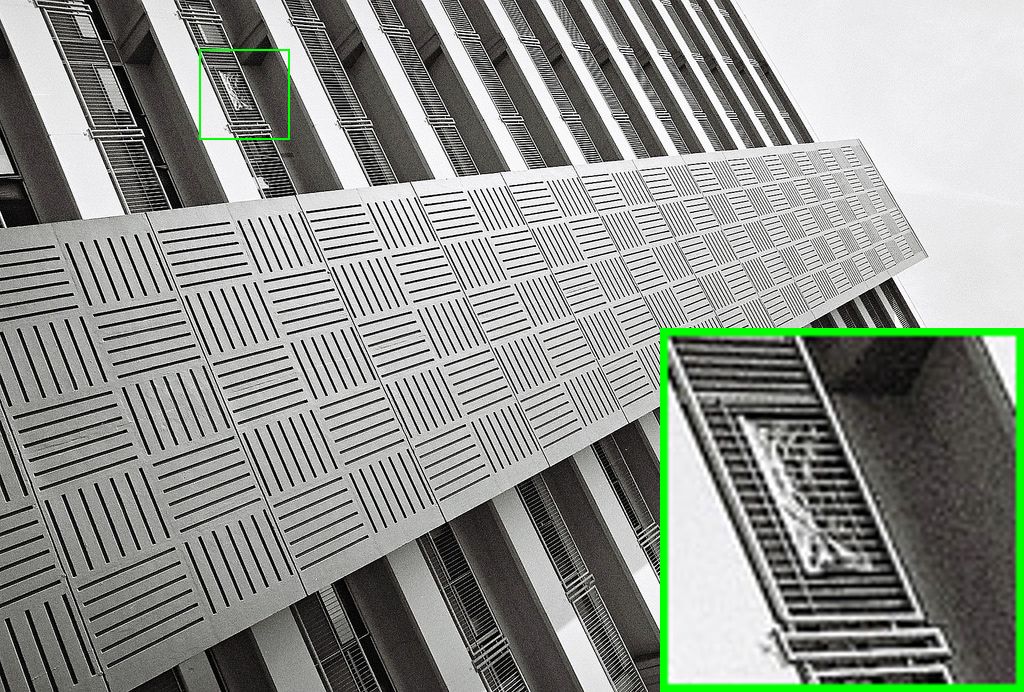} }
{\footnotesize  PSNR/SSIM \\ (a) HR }
\end{minipage}
\begin{minipage}[t]{0.15\textwidth}
\centering 
\raisebox{-0.15cm}{\includegraphics[width=1\textwidth]{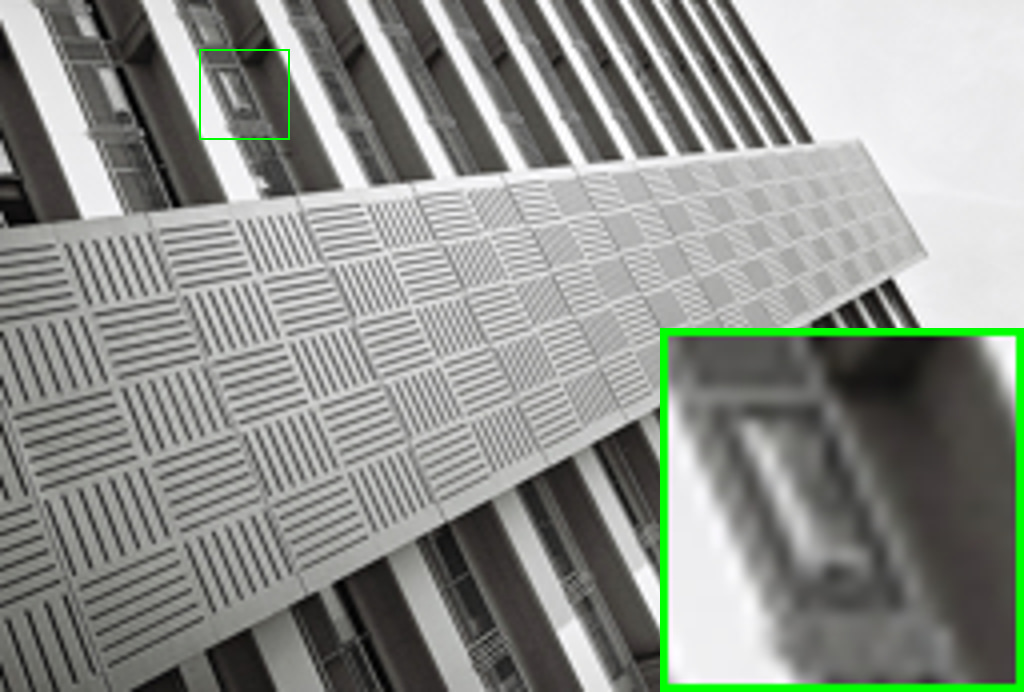}}
{\footnotesize  16.61 dB/0.4398  \\ (b) LR bicubic}
\end{minipage}
\begin{minipage}[t]{0.15\textwidth}
\centering 
\raisebox{-0.15cm}{\includegraphics[width=1\textwidth]{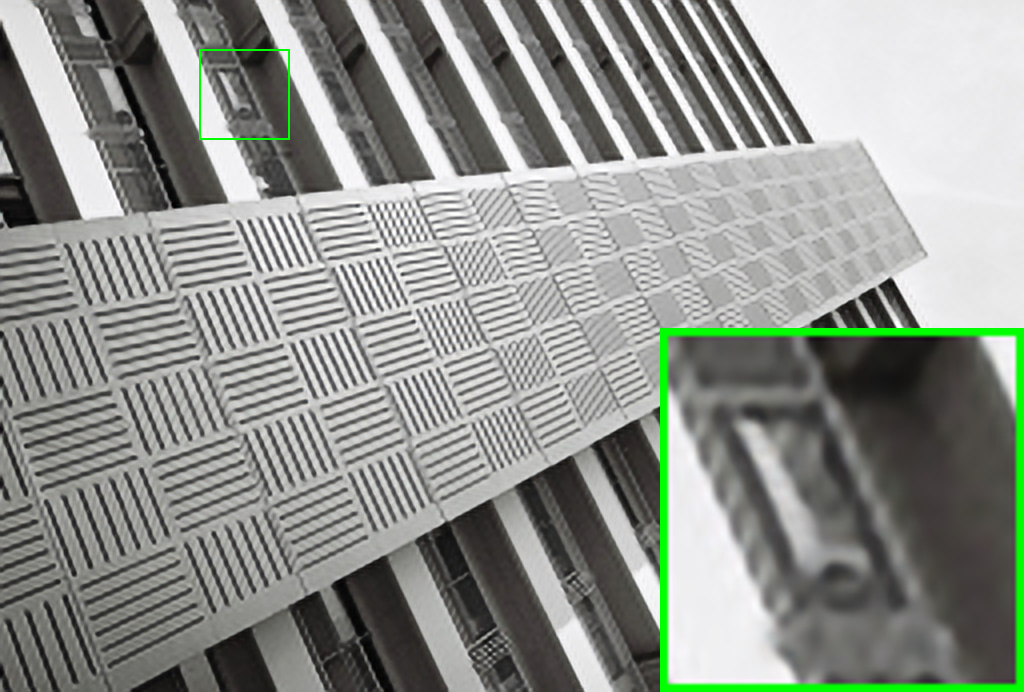}}
{\footnotesize  17.15 dB/0.5061 \\ (c) SRCNN~\cite{dong2015image} }
\end{minipage}
\begin{minipage}[t]{0.15\textwidth}
\centering 
\raisebox{-0.15cm}{\includegraphics[width=1\textwidth]{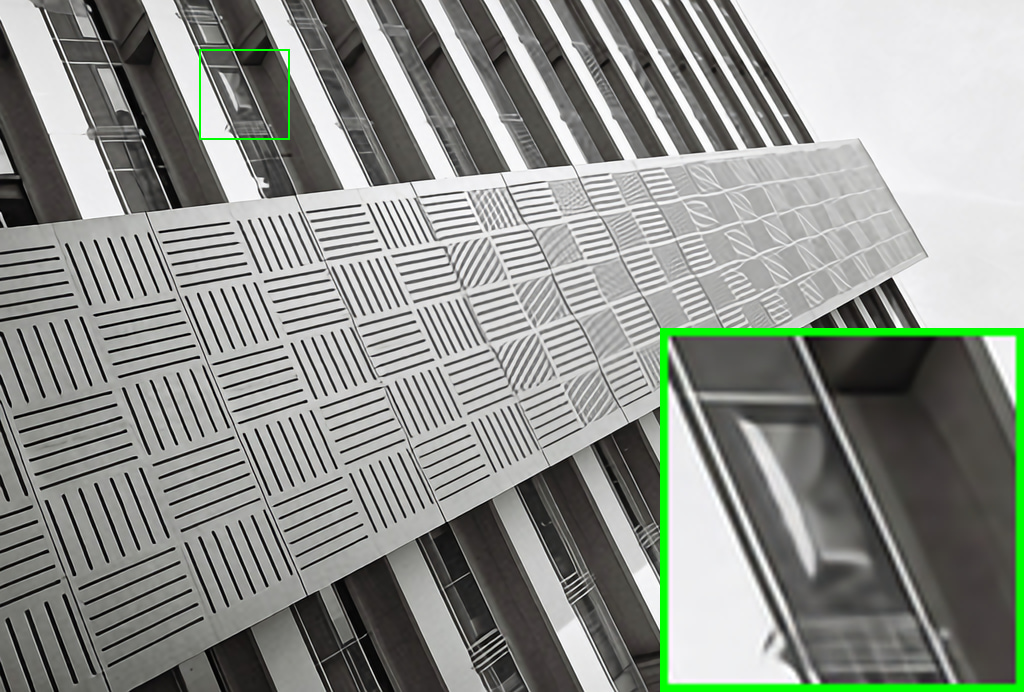}}
{\footnotesize  19.15 dB/0.6785 \\ (d) EDSR~\cite{lim2017enhanced}}
\end{minipage}
\begin{minipage}[t]{0.15\textwidth}
\centering 
\raisebox{-0.15cm}{\includegraphics[width=1\textwidth]{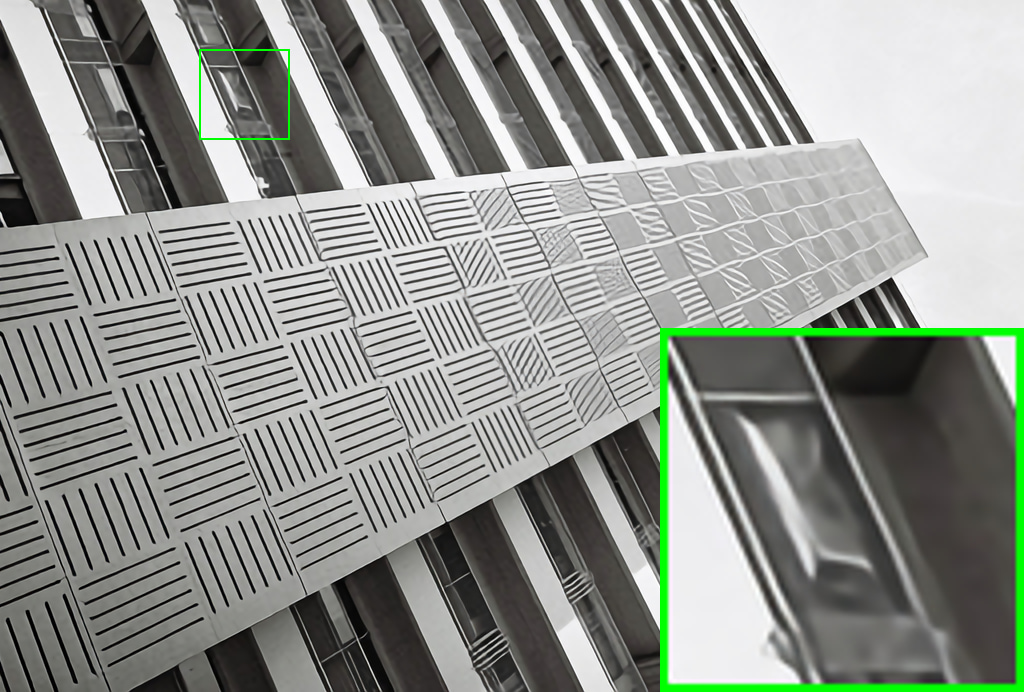}}
{\footnotesize  19.08 dB/0.6682 \\ (e) MSRResNet~\cite{DBLP:conf/eccv/ZhangDLTLTWZHXL20}}
\end{minipage}
\begin{minipage}[t]{0.15\textwidth}
\centering 
\raisebox{-0.15cm}{\includegraphics[width=1\textwidth]{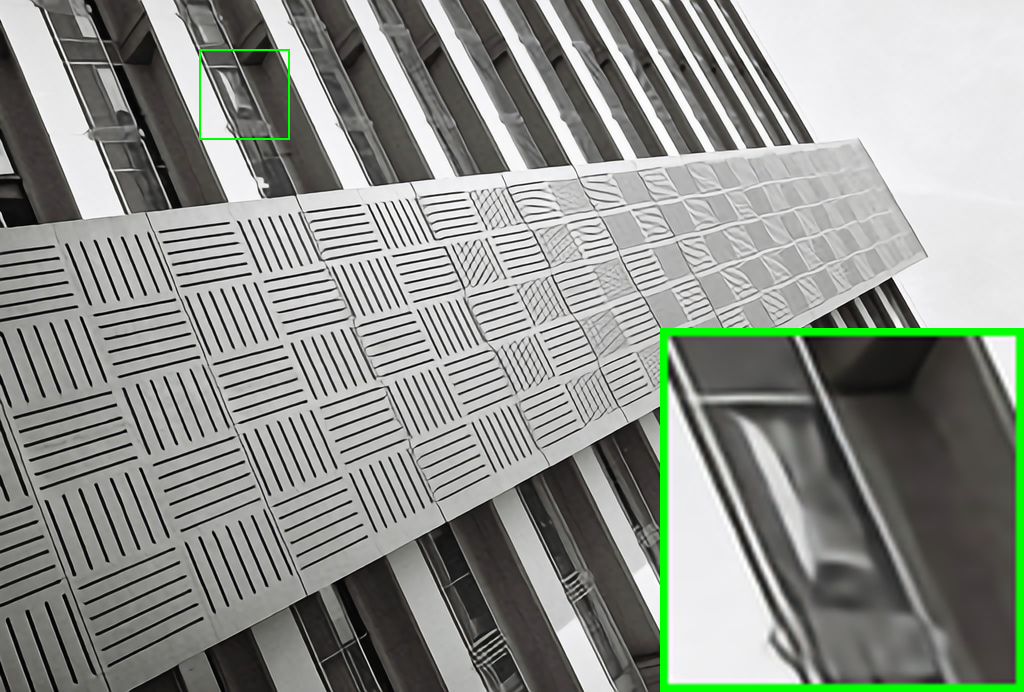}}
{\footnotesize   19.39 dB/0.6830 \\ (f) MSRResNet+PFCA}
\end{minipage}
}
\caption{\textbf{Visual quality and PSNR/SSIM results} by different image super-resolution methods on Set5 (1st row), Set14 (2nd row) and Urban100 (3rd row).}
\label{sr-imgshow}
\vspace{-3mm}
\end{figure*}

\section{Experiments}
\label{sec:experiment}

\subsection{Experiment Details}

In our experiments, image classification task is performed on ImageNet 2012 dataset~\cite{deng2009imagenet} and CIFAR-100 dataset~\cite{krizhevsky2009learning}. ImageNet 2012 dataset contains 1.28 million train images and 50,000 valid images with total 1000 classes. CIFAR-100 dataset contains 50,000 train images and 10,000 test images with 100 classes. Image super-resolution task is performed on DIV2K dataset~\cite{Agustsson_2017_CVPR_Workshops} with contains 800 images to train. Experiments are performed by using Pytorch framework with one NVIDIA RTX 3090 GPU.

\subsection{Image Classification}

When training classification networks on ImageNet dataset, input images are adjusted to size $224\times224$. We use the same Data augmentation method in \cite{he2016deep,hu2018squeeze}. For optimization process, SGD optimizer is used with momentum 0.9 and a mini-batch size 128. The learning rate is initially set to 0.1 and reduces to $0.1$ of the previous value for every 30 epochs. Weight decay is used and set to $10^{-4}$.
The network models are trained totally 100 epochs from scratch.
Classification test result is presented on the Table~\ref{table-imagenet}. Flops are computed with a single image with size $224\times224$. It is indicated that adding PFCA module could effectively increase the classification accuracy on both top-1 and top-5.



We follow the same optimization strategy in \cite{krizhevsky2009learning} to train ResNet on CIFAR dataset where batch size is 128 and the model trained for 200 epochs. The initial learning rate here is set to 0.05 and divided by 5 at epoch 60, 120, and 160 and weight decay is $5\times10^{-4}$. L1 loss function is used to train the network and SGD optimization strategy is used with momentum set to 0.9. We present top-1 and top-5 accuracy as test metrics on both datasets. Classification test result is presented in the Table~\ref{table-imagenet}. It is indicated that adding PFCA module could effectively increase the classification accuracy.
One can see that our PFCA module also helps increase the classification accuracy on CIFAR-100 dataset.

\subsection{Image Super-Resolution}

We further testify the effectiveness of our proposed Parameter-Free Channel Attention module on image super-resolution task. To effectively compare different attention, we provide four candidate models to train: the baseline model MSRResNet~\cite{DBLP:conf/eccv/ZhangDLTLTWZHXL20} with no attention, with PA~\cite{zhao2020efficient}, with CA and with PFCA. 
The number of residual blocks is 16. We train models on DIV2K train dataset~\cite{Agustsson_2017_CVPR_Workshops} which contains 800 high resolution images.
The HR images are randomly cropped with patch size $256\times 256$. Mini-batch size is set to 32. The cosine annealing learning scheme is used when training model, and learning rate is set with $2\times10^{-4}$ initially and the minimum value to $10^{-7}$. We set weight decay to $5\times10^{-8}$. Restart period is set after each 250k iterations. All models are trained for total 1000k iterations.
The corresponding LR images are obtained by downsampling HR images by 4 downscale factor using bicubic method. After the training stage, we use five standard benchmark datasets, Set5~\cite{bevilacqua2012low}, Set14~\cite{zeyde2010single}, B100~\cite{martin2001database}, Urban100~\cite{huang2015single} and Manga109~\cite{matsui2017sketch}, to test the performance. Evaluation metric is set with average peak signal to noise ratio (PSNR) and average structural similarity index (SSIM) on each test set. All test metrics are calculated on the Y channel of images. 
In addition to the comparison of MSRResNet~\cite{DBLP:conf/eccv/ZhangDLTLTWZHXL20} and our model, we also compare with SRCNN~\cite{dong2015image} and EDSR~\cite{lim2017enhanced} results in the paper, since the former is the first CNN model for super-resolution task and the latter one has similar architecture as ours but with much more parameters.

\noindent
\textbf{Objective results} are provided in Table~\ref{table-sr-evaluation}.
We observe that adding PFCA in the residual block helps improve the PSNR and SSIM values compared to the baseline MSRResNet.
Adding different attentions could bring effect improvement but with different parameters.
Note that the PSNR/SSIM values of PFCA are similar to those of CA.
That means PFCA could achieve comparable performance of CA with little parameter growth.

\noindent
\textbf{Comparisons on visual quality} are presented in Figure~\ref{sr-imgshow}. Images are chosen from the test set. The visual result and their corresponding test metrics are listed below the image. It reveals that the restored quality of image of adding PFCA part is better in the local texture part. Thus the PFCA module could effectively help improve the visual quality.

\section{Conclusion}
\label{sec:conclusion}

In this paper, we proposed a Parameter-Free Channel Attention (PFCA) that could serve the function of channel importance estimation, while not introducing additional parameters. We put this PFCA module into the image classification model as well as the image super-resolution model and then performed experiments on image classification and image super-resolution. The experiment results validate that our PFCA module could function as an effective attention module in deep networks. According to some experiment results, the fixed structure of PFCA needs to be further improved to achieve better performance compared to CA.




\bibliographystyle{IEEEbib}
\bibliography{refs}

\end{document}